\documentclass{aa}  
\usepackage{graphicx}
\usepackage{lscape}
\usepackage{hyperref}
\usepackage{txfonts}
\usepackage{float}
\begin{document}

\title{Identification of quasars variable over long time scales 
from infrared surveys.}
\subtitle{Ensemble variability and structure function properties}

\author{E. A. Zaharieva,\inst{1}
        V. D. Ivanov,\inst{2}
        E. P. Ovcharov,\inst{1}
        O. I. Stanchev \inst{1}
        }

   \institute{Department of Astronomy, Faculty of Physics, University of Sofia, BG-1164 Sofia, Bulgaria\\
              \email{ezaharieva@phys.uni-sofia.bg}
         \and
             European Southern Observatory, Karl-Schwarzschild-Str. 2, D-85748 Garching bei München, Germany\\
             \email{vivanov@eso.org}
             }

   \date{Received June 03, 2024; accepted April 22, 2025}

 
\abstract
{Quasars are variable and their variability can both constrain their 
physical properties and help to identify them. 

}
{We look for ways to efficiently identify quasars exhibiting 
consistent variability over multi-year {time-scales}, based on {a} small 
number of epochs. Using {infrared} (IR) is desirable to avoid {bias}
 against reddened objects.

}
{We compare the apparent brightness of known quasars that have been 
observed with two IR surveys, covering up to {a twenty-year} baseline: 
the Two Micron All Sky Survey (2MASS; 1997-2001) and the VISTA 
Hemisphere Survey (VHS; 2009-2017). We look at the previous studies 
of the selected variable quasars to see if their variable behaviour 
is known and when available we use {multi-epoch} monitoring with 
the Zwicky Transient Facility (ZTF) to obtain a measure of optical 
variability of individual objects.
}
{We build a sample of nearly 2500 quasars that show statistically 
significant variability between the 2MASS and VHS. About 1500 of
these come from the new Quaia sample based on {\it Gaia} spectra 
and about 1/3 of these have hardly been studied. The Quaia sample 
constitutes the main product of this work. Based on ensemble 
variability and structure function analysis we demonstrate that 
the selected objects in our sample are representative of the 
typical quasar population and show behaviour, consistent with 
other quasar samples. Our analysis strengthens previous results, 
for example that {variability} decreases with the rest-frame wavelength and 
that it exhibits peaks for certain absolute magnitudes of the 
quasars. Similarly, the structure function shows an increase in 
{variability} for rest-frame time lags {below} $\sim$1500\,d and {a}
decrease for longer lags, just like in previous studies.
}
{Our selection, even though it is based on two epochs only, {seems} 
to be surprisingly robust, showing up to $\sim$11\,\% contamination 
by quasars that show stable non-variable behaviour in ZTF.
}

\titlerunning{Quasars with long-term variability}
\authorrunning{Zaharieva et al.}
\maketitle
%

\section{Introduction}

Quasi-stellar objects {(QSO)} or quasars are powerful and distant 
active galactic nuclei {(AGN)} that were identified in {the} 
early 1960s 
\citep{1963Natur.197.1041G,1963Natur.197.1040O,1963Natur.197.1040S}. 
Some of the early quasars were well known to vary on short time scales 
which implied a compact structure including a central continuum source 
surrounded by emission-line {region} 
\citep{1963ApJ...138...30M,1963Natur.198..650S}. 
Soon accreting super-massive black holes were proposed as quasars' 
energy sources 
\citep{1963Natur.197..533H,1963MNRAS.125..169H,1964ApJ...140..796S}. 
Subsequent monitoring demonstrated that the compact nature of the 
black holes agreed well with the observed variability timescales 
\citep{1961PASP...73..292S,1963Natur.198..650S,1973AJ.....78..353A,1982ApJ...255..419B,1994MNRAS.268..305H,1994ApJ...433..494T}.
\citet{1985ApJ...297..621A} and \citet{1993ARA&A..31..473A} 
organized the rich observational results of the preceding decades 
into the unified model of active galaxies that is commonly accepted 
{today.}
  
{ The variability} proved {to be} a useful tool in at least 
two aspects -- 
to put observational constraints on the spatial structure of their 
{inner} regions and {to} measure the masses of the central black holes 
\citep{1982ApJ...255..419B,1993PASP..105..247P,2004ApJ...613..682P},  
and to help identify QSOs from their variability patterns in large 
synoptic sky surveys 
\citep{1990Msngr..61...46H,2008A&A...488...73T,2010ApJ...714.1194S}. 
\citet{2021IAUS..356..101L} summarized the recent understanding of 
QSO variability studies. 

{  
Previous optical long-term variability studies 
\citep[e.g.,][]{2003AJ....126.1217D,2005AJ....129..615D} indicated 
that:
\begin{itemize}
\item quasars are more variable toward the blue than {toward} the red;
\item variability increases {monotonically} with increasing time lag;
\item less luminous quasars are more variable than {more luminous
ones};
\item blazars tend to be more variable than other quasars.
\end{itemize} 
}

Various variability mechanisms have been discussed: thermal 
instabilities in the accretion disk \citep{1976MNRAS.175..613S},
circumnuclear starburst or accretion instability 
\citep{1998ApJ...504..671K}, {and} interaction between a complex 
multi-structured accretion disk \citep{1991ApJ...380L..51H}, among 
others.

{  
The need to create complete samples of variable quasars became 
obvious early on \citep[e.g.,][]{1983LIACo..24...31H}, both because 
they would give a {glimpse of quasars as a population} and because 
{large samples} would hopefully include rare objects that have 
always been useful in identifying and characterizing the most extreme
physical processes responsible for the variability. 

Reliable detected variable quasar samples have multiple applications.
One of the most obvious is finding binary black holes -- a process 
that gives us a direct glimpse into the late stages of galaxy merging 
and the growth of {supermassive} black holes and the bulges that 
they are immersed in. \cite{2020MNRAS.499.2245C} reported a 20-{year} 
optical monitoring. Their light curves demonstrate the two major
obstacles {to} this goal: small amplitudes, compared with the 
observational uncertainties, and long {time-scales} that can reach many 
years \citep[see also][]{2021MNRAS.508.2937M}.

Next, the variability helps to constrain the processes in the 
accretion disks surrounding {supermassive} black holes. The 
classical accretion disk models successfully explained the AGN 
dichotomy \citep{1973A&A....24..337S,1993ARA&A..31..473A}, but a 
more careful look, made possible by the wealth of recent variability
and spectral energy distribution data indicates {problems} that
\citet{2013Natur.495..165A} listed some: origin of the radio emission,
strength of the ultraviolet flux, properties of the innermost disk, 
as constrained from the X-ray observations. {In the context of 
this work,} unclear changes in the structure of the accretion disk. 
Theorists resort to new ideas, such as opacity variations, to
explain the observed luminosity changes with changes in the disk 
scale height \citep[][]{2020ApJ...900...25J}.

Most recently, \citet{2023MNRAS.524..188L} demonstrated that {5 to 7-year}-
long optical and mid-infrared monitoring can help to identify such 
rare objects as changing look AGNs. Furthermore, 
\citet{2024arXiv241022671A} selected {changing-look} candidates from 
{optical light curves alone}.

}

{Motivated by these considerations and the growing number of newly identified
quasars from ongoing and future surveys, we revisit the topic of quasar variability.
In particular, we identify new, extremely variable quasars that can serve as
valuable targets for studying the spatial structure and circumnuclear physical
conditions of quasars in the era of deep, large-scale sky surveys, including massive
spectroscopic efforts.} A closer look at the nearby active galaxies 
indicated that their nuclear regions contain dust that may obscure 
the broad lines, to the point of making them inaccessible in the 
optical \citep{1995ApJ...454...95M,2000ApJ...531..219M}. To alleviate 
this issue we look at the variability in the infrared (IR) spectral 
{region.} Today it is 
established that the IR emission of active nuclei is {thermal in} 
origin, coming from circumnuclear dust, heated up by the nuclear 
source \citep{1981ApJ...250...87R}; non-thermal IR emission would 
imply no time delay between UV/optical and IR, which contradicts 
the results from the multi-wavelength monitoring campaigns 
\citep{1993ApJ...409..139S}. The IR variability is usually 
combined with the optical, and helps to map the dust emission 
region \citep[e.g.,][]{2011A&A...534A.121H}. 

Here we report two lists of variable quasars. Their follow-up, 
for example along the line of the periodicity analysis of
\citet{2021MNRAS.508.2937M} and Minev et al. (submitted) will be 
reported elsewhere. 
The next section describes the motivation for this work, 
Sec.\,\ref{sec:samples} {outlines} the photometric surveys, the sources 
of the initial quasar samples, and the identification of the most 
highly variable quasars. Sections\,\ref{sec:ensable_varisb} and 
\ref{sec:sf} present the ensemble variability and the structure 
function parts of the study, respectively. Finally, in 
Sec.\,\ref{sec:discussion} we discuss the nature and properties of 
the quasars in other wavelengths and Sec.\,\ref{sec:conclusions}
summarizes this work.

\section{Motivation and outlook}\label{sec:motivation}

We characterize the long-term variability of a set of active galaxies 
in the Southern Hemisphere, in preparation for the future facilities 
that will be able to deliver spectroscopic follow-ups for thousands of 
objects: 4MOST \citep[4-metre Multi-Object Spectroscopic 
Telescope;][]{2019Msngr.175....3D}, MOONS \citep[Multi-Object Optical 
and Near- Spectrograph;][]{2020Msngr.180...10C} and perhaps 
the future WST \citep[Wide‑field Spectroscopic survey 
Telescope;][]{2023AN....34430117B}.
The challenge in the target selection for these projects is the need 
to assemble a set of highly variable objects, that are likely to 
maintain this variability level over long {time-scales}, 
{guaranteeing} -- as much as possible -- that the investment of valuable 
spectroscopic time over  many years will yield meaningful information about the monitored 
objects. The LSST eventually will help to build such a sample, with two 
caveats. First, it saturates at 
$r$$\sim$15.8\,mag\footnote{\href{https://www.lsst.org/sites/default/files/docs/sciencebook/SB_3.pdf}{https://www.lsst.org/sites/default/files/docs/sciencebook/SB\_3.pdf}}, 
making the spectroscopic follow-up of large samples time-consuming. 
Second, the selection is optical, biasing the sample somewhat against 
obscured -- intrinsically or not -- objects. 

Other teams have already worked toward selecting a sample of variable 
objects. For example, \citet{2023A&A...675A.195S} carried out a
sophisticated hierarchical balanced random forest {modelling}, on a 
combination between mid- and optical colors, optical morphology, and 
4-year-long optical monitoring from ZTF \citep[Zwicky Transient 
Facility;][]{2019PASP..131a8002B}.

Here, we adopt a different and {perhaps} somewhat more robust approach. 
First, we perform our selection on a much longer baseline, in the hope 
of ensuring {consistently} variable behavior over a longer time 
scale: we combine data from The Two Micron All Sky Survey 
\citep[2MASS;][]{2006AJ....131.1163S} and The VISTA Hemisphere Survey 
\citep[VHS;][]{2013Msngr.154...35M}, covering about two decades. This 
also means that our selection is based on the {IR} wavelength region, 
alleviating the bias against redder and potentially more obscured 
objects. Another important consideration for selecting the VHS was 
that most {future} facilities will be located in the South, 
making the new targets accessible to them.

Furthermore, unlike \citet{2023A&A...675A.195S}, our starting point 
is a sample of spectroscopically confirmed quasars, making it much
more efficient, because we do not have to search for and to confirm 
the quasars, their nature is already securely known. We begin with 
the well-studied sample of {QSO and AGN} catalogue 
of \citep[][13th Edition; VCV10, hereafter]{2010A&A...518A..10V} to 
ensure that we do identify extremely variable quasars. Next, we take 
advantage of the newly reported list of quasars selected from {\it 
Gaia} based on its low-resolution BP/RP spectra to build a new and 
larger sample of highly variable quasars in the Southern hemisphere 
\citep{2024ApJ...964...69S}. We also examined whether our selection 
identifies highly variable quasars by closely following the example 
of \citet{2012ApJ...747...14K} and verified with a literature survey 
that this approach is successful. To achieve this goal, we perform 
the usual structure function analysis. 

{We compare the properties of objects in our sample with those of
\citet{2012ApJ...747...14K}, because the IR data that we use are 
similar, and an agreement would land credibility to our results and
selection. Their work and our work are complementary, covering both 
the Northern and the Southern sky, and together we can potentially 
yield a reasonably uniform sample of highly variable quasars and 
AGNs spanning the entire sky. However, we concentrate first on the 
Southern sample, because it is potentially less explored -- e.g., 
there is no Souther analog of ZTF -- and defining it is more urgent 
in the context of the {\it Rubin} observatory that will soon begin
operations.}


\section{Sample selection and survey data}\label{sec:samples}

{ 
Here the infrared variability of quasars is estimated as a simple 
difference between their 2MASS and VHS apparent magnitudes (corrected
for the {colour} differences as described in the Sec.\,\ref{sec:corr}).
Multiple measurements are available from VHS for some objects; we 
only used one, selected at random, and ignored the rest, because the 
time span of the VHS is {typically} shorter than the baseline that the 
2MASS-VHS comparison provides and our goal is to reliably select 
objects that do show long-term variability. Therefore, the number of 
measurements in our infrared light curves is always two.
}

\subsection{Survey data and initial quasar samples}

The 2MASS is a nearly all-sky imaging survey in $J$ (1.25 $\mu$m), 
$H$ (1.65 $\mu$m), and $K_S$ (2.16 $\mu$m; denoted from now on as 
$K$ for simplicity) conducted between {June} 1997 and February 2001. 
Observations were obtained at two {1.3-metre} telescopes positioned 
at Mount Hopkins, Arizona, USA, and Cerro Tololo, Chile, achieving
signal-to-noise S/N=10 at $J$=15.8\,mag and $K$=14.3\,mag.

The VHS footprint spans 16,730 deg of the southern sky in $Y$ (1.02 
$\mu$m), $J$ (1.25 $\mu$m), $H$ (1.65 $\mu$m) and $K$ (2.15 $\mu$m) 
bands, conducted between November 2009 and March 2017. Observations 
were obtained at {4.1-metre} Visible and Survey Telescope for Astronomy 
\citep[VSTA;][]{2006Msngr.126...41E} with the VISTA  camera 
\citep[VIRCAM;][]{2006SPIE.6269E..0XD}, reaching a 5$\sigma$ median 
point source detection at $J_{AB}$=20.8\,mag and $K_{AB}$=20.0\,mag, 
with a minimal exposure time of 60 seconds. 

These two surveys have three overlapping bands, but we consider this 
redundant for our purposes, and in further analysis we only use $J$ 
and $K$, which is also sufficient to directly compare our results 
with \citet{2012ApJ...747...14K}, who performed a similar analysis 
earlier on different IR data sets. Two bands {are} the minimum to provide 
a cross-check of individual objects. We consider other filters and 
different parts of the electromagnetic spectrum when investigating 
the behaviour of objects of interest, e.g., if they show extreme 
variability, or {contradictory} results in the selected two bands.

\begin{figure}[h!]
\centering
\includegraphics[width=0.99\columnwidth]{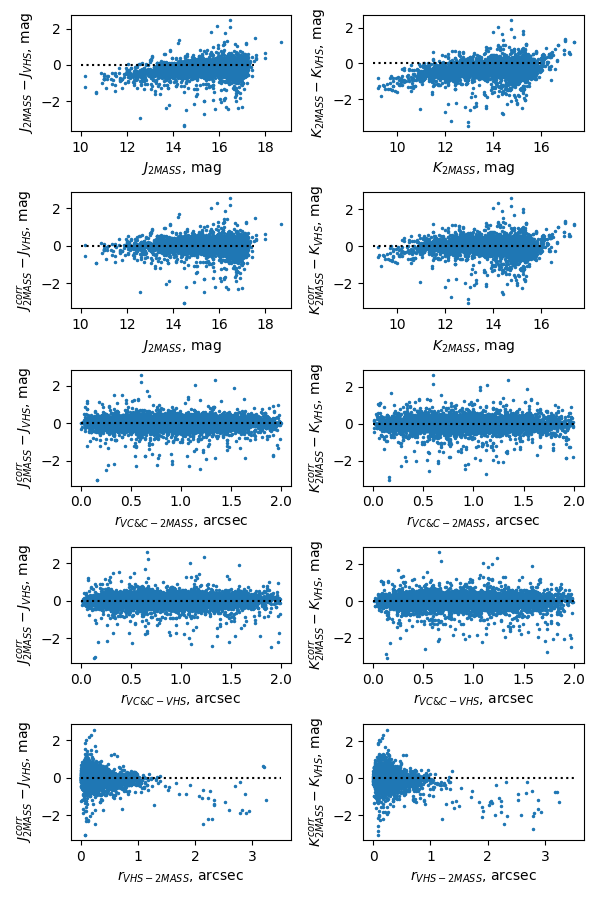}
\caption{Matching the 2MASS and VHS sources. 
The conversion of the 2MASS to the VHS photometric system is 
illustrated on the first two rows: the panels on the top row show 
the magnitude differences for the sample quasars before the 
correction and on the second -- after the conversion according 
to Eq.\,\ref{eg:homog} and Table\,\ref{tab:corr_coeff}. The 
deviation at sources brighter than $\sim$11\,mag is related to the 
saturation in VHS.
The astrometric properties are illustrated on the remaining three 
rows, showing the magnitude differences as a function of the angular 
separation between VCV10 and 2MASS coordinates, between VCV10 and VHS 
coordinates and between the 2MASS and VHS coordinates (top to 
bottom). 
Panels on the left are for $J$ and on the right for the $K$ band.
}\label{fig:Corr_crit}
\end{figure}

Our bright reference {QSO and AGN} sample is based on {VCV10}. 
It contains 133\,336 quasars, 1\,374 BL Lacs and 34\,231 active galaxies 
(including 16\,517 Seyfert 1s). 
We selected a subset 
of quasars with photometric data in both the 2MASS and VHS, adopting 
a liberal search {radius} around the VCV10 positions of 2\arcsec\ 
(Fig.\,\ref{fig:Corr_crit}, bottom three rows of panels). The 
average offset with 2MASS was 0.9$\pm$0.4\arcsec. The average 
separation of the matching 2MASS and VHS sources was 
0.3$\pm$0.3\arcsec and only $\sim$0.8\,\% of the objects could be
attributed to a tail-like structure at $>$1.6\arcsec which is not
surprising, because the VISTA public surveys are astrometrically 
calibrated with 2MASS. 

The magnitude differences between counterparts selected from the two 
surveys show no trends with angular separation, hinting that perhaps 
the VCV10 coordinates may not be very accurate -- {which is not surprising} 
because in many cases they come from older literature based on data 
with relatively poor spatial resolution, and maybe are affected by 
the contribution of the underlying host galaxy. To verify our 
selection we also compared {the} positions of the 
counterparts in 2MASS and VHS (bottom two panels), and found that 
the vast majority of sources are located closer than 1.2\arcsec\ 
apart. This criterion was met by 7\,019 objects for $J$ band and 
6\,721 for $K$. 

The Quaia catalogue of \citet{2024ApJ...964...69S} presents a new 
all-sky addition to the quasar census, valuable with its high 
fidelity \citep{2024A&A...687A..16I}. It is based on the Gaia 
low-resolution BR/RP spectra and unWISE mid-{colours} 
\cite{2019ApJS..240...30S}. The final Quaia list contains about 
1.3 million objects down to $G$$<$20.5\,mag. A cross-correlation 
with VHS yielded about 550\,000 matching sources within 
$\sim$1.1\arcsec, but close to 5.8 million {measurements} -- this 
includes all filters and in some cases multiple measurements in 
the same filter; they extend over {many year long intervals}. 
Next, we cross-{correlated} this list with 2MASS to obtain quasars 
with nearly {two decades of} photometric coverage. This left a sample 
of 319\,853 objects or 245\,930 in only $J$ and $K$ band VHS 
observations are considered.

\subsection{Sample curation and variability detection}\label{sec:corr}

The filter transmission curves and the detector sensitivity curves 
of 2MASS and VHS differ, albeit slightly, leading to systematic 
differences between the magnitudes in the two catalogues. 
Fig.\,\ref{fig:Corr_crit} (top panels) shows this for the VCV10 
sample: the deviations are greater for the brighter objects, probably 
due to non-linearity effects at the image core pixels, and because of 
systematic colour {differences} between bright and faint objects. To 
{homogenise} the data, we introduced a correction, converting the 2MASS 
magnitudes to VHS magnitudes:
\begin{equation}\label{eg:homog}
m_{\rm VHS} = m_{\rm 2MASS} \times a + b
\end{equation}
where $a$ and $b$ are the fitting coefficients listed in 
Table\,\ref{tab:corr_coeff}. The effect of the correction is shown 
in Fig.\,\ref{fig:Corr_crit} (second top panels). There is an apparent 
deviation at the bright magnitudes, probably because the pixels in 
{the} image cores of the brightest {objects} saturate. We 
adopt a cautious 
approach, omitting from our analysis objects brighter than {11\,mag 
in any band. This condition was applied to the final samples, some 
plots (e.g. Fig.\,\ref{fig:Corr_crit}) contain brighter objects to 
demonstrate the effect of the non-linearity.}

\begin{table} 
\caption{\label{tab:corr_coeff}Conversion of the VHS magnitudes to
the 2MASS system.} 
\begin{center}
\begin{tabular}{ccc} 
\hline
\noalign{\smallskip}
Filter & a & b \\
\noalign{\smallskip}
\hline
\noalign{\smallskip}
$J$ & 0.9153$\pm$0.0022 & 1.522$\pm$0.035 \\
$K$ & 0.9037$\pm$0.0034 & 1.649$\pm$0.049 \\
\noalign{\smallskip}
\hline
\end{tabular}
\end{center}
\end{table}

\begin{table*}[h!]
\caption{List of {952} objects, selected from the VCV10 sample, 
with photometry from 2MASS and VHS surveys, that show statistically 
significant variability in both $J$ and $K$ bands. Only a fraction 
is shown {for} guidance; the complete table is available in the 
electronic edition of the journal.}\label{tab:var_sample_VCV10} 
\tiny
\begin{center}
\begin{tabular}{@{}c@{ }c@{ }c@{ }c@{ }c@{ }c@{ }c@{ }c@{ }c@{}} 
\hline
RA DEC (J2000) & 2MASS ID & $J^{2MASS}$ & $K^{2MASS}$ & $J^{VHS}$ & $K^{VHS}$ & ~S/N($\Delta~J$)~ & ~S/N($\Delta~K$)~ & ~S/N($\Delta~J$,$K$)\\
\hline
12:56:11.1 $-$05:47:21 &~~J12561117$-$0547215~~~& 12.585 0.027 & 10.941 0.026 & 15.490 0.006 & 13.489 0.004 & 96.31 & 81.81 & 89.06 \\
02:10:46.3 $-$51:01:02 & J02104620$-$5101018 & 13.692 0.026 & 11.977 0.024 & 16.071 0.007 & 14.694 0.008 & 81.65 & 95.64 & 88.64 \\
22:35:13.3 $-$48:35:59 & J22351322$-$4835588 & 13.809 0.029 & 12.124 0.026 & 16.026 0.006 & 14.406 0.006 & 68.53 & 73.95 & 71.24 \\
\multicolumn{9}{l}{...} \\
01:05:26.4 $-$00:48:06 & J01052630$-$0048066 & 15.786 0.111 & 14.449 0.126 & 16.291 0.008 & 15.068 0.012 &  3.14 &  3.16 &  3.15 \\
02:47:07.6 $-$07:28:02 & J02470759$-$0728022 & 16.843 0.164 & 15.347 0.181 & 17.393 0.014 & 16.037 0.027 &  3.01 &  3.13 &  3.07 \\
22:52:50.8 $-$24:27:37 & J22525076$-$2427378 & 16.921 0.166 & 15.745 0.288 & 17.471 0.018 & 16.702 0.052 &  3.02 &  3.11 &  3.06 \\
\noalign{\smallskip}
\hline
\end{tabular}
\end{center}
\tablefoot{The columns contain position, 2MASS ID and magnitudes, 
VHS magnitudes, {and} statistical significances of the 2MASS vs. VHS 
magnitude differences for $J$ and $K$ separately, {as well as the} average 
(the objects are sorted by the latter).}
\end{table*}

\begin{figure*}[h!]
\centering
\includegraphics[width=1.7\columnwidth]{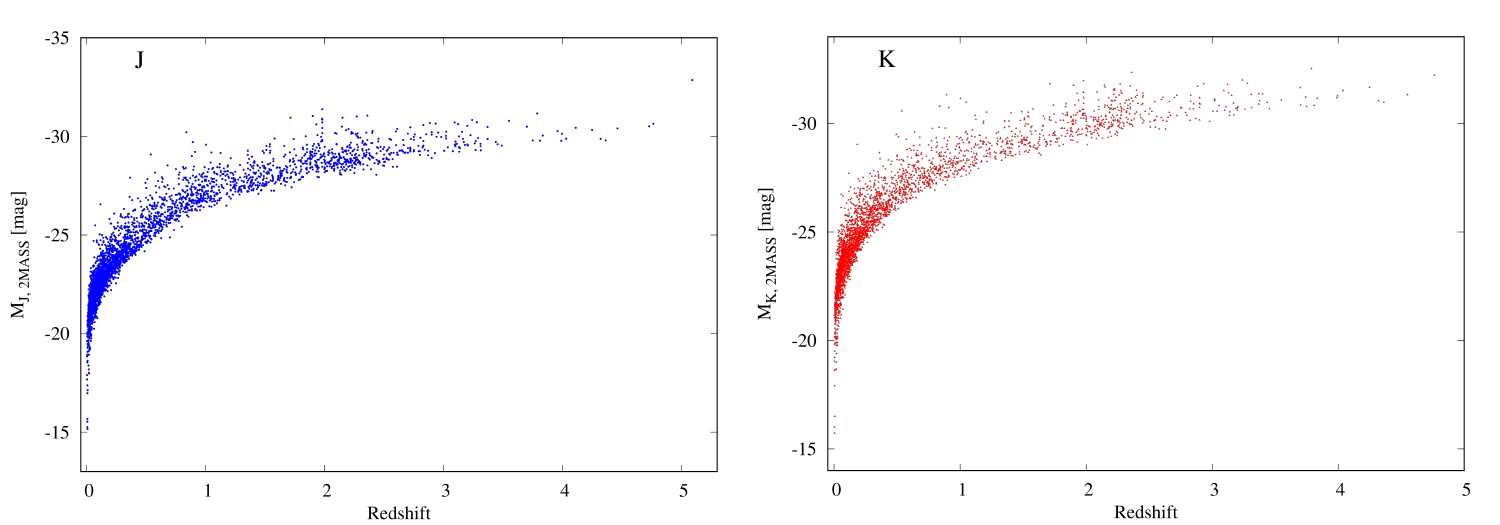}
\includegraphics[width=1.75\columnwidth]{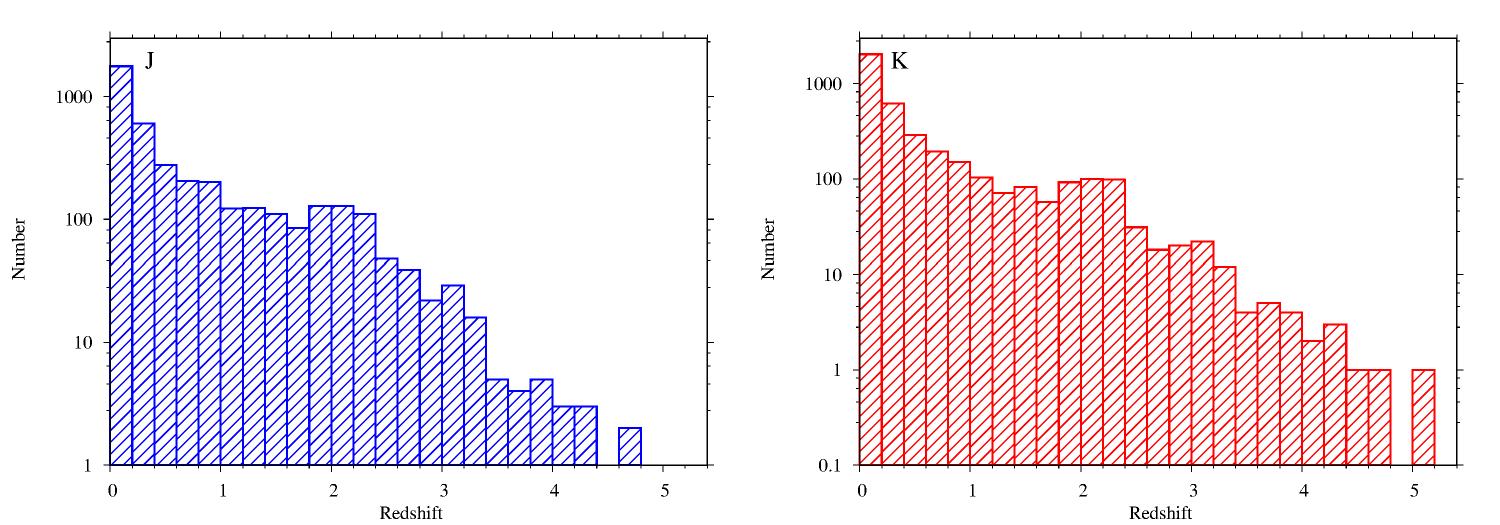}
\caption{Properties of the VCV10 sample. 
{\it Top:} {  Absolute magnitudes in $J$ and $K$ versus redshift, 
respectively.} 
{\it Bottom:} {  Histograms of the redshifts of quasars in the sample
for $J$ and $K$ bands, respectively.
They span redshifts up to $z$$\sim$5.0, but {approximately} 2/3 have $z$$\leq$0.5 
and only {about} $\sim$10\,\% have $z$$>$2.0.}}
\label{M_z_J_K}
\end{figure*}

Overall, the sample spans a redshift range from 0.003 to 5.1.
Fig.\,\ref{M_z_J_K} shows the absolute magnitude as a function of 
redshift. The absolute magnitudes were determined assuming 
H$_0$ = 73.04\,km\,$s^{-1}$$ Mpc^{-1}$, $\Omega_M$ = 0.3089, 
and $\Omega_\Lambda$ = 0.6911. We applied K-corrections based on 
a power-law spectrum with $\alpha$ = 0.5:
\begin{equation}
K(z) = 2.5*(\alpha - 1) * log(1+z)
\end{equation}

Finally, having the corrected magnitudes allows us to identify the 
{AGNs and quasars} that show variability at 3-$\sigma$ level from:
\begin{equation}
|\Delta m_i| > 3 \times \sqrt{\sigma_{i, {\rm 2MASS}}^2+\sigma_{i, {\rm VHS}}^2}
\end{equation} 

\begin{figure}[h!]
\centering
\includegraphics[width=\hsize]{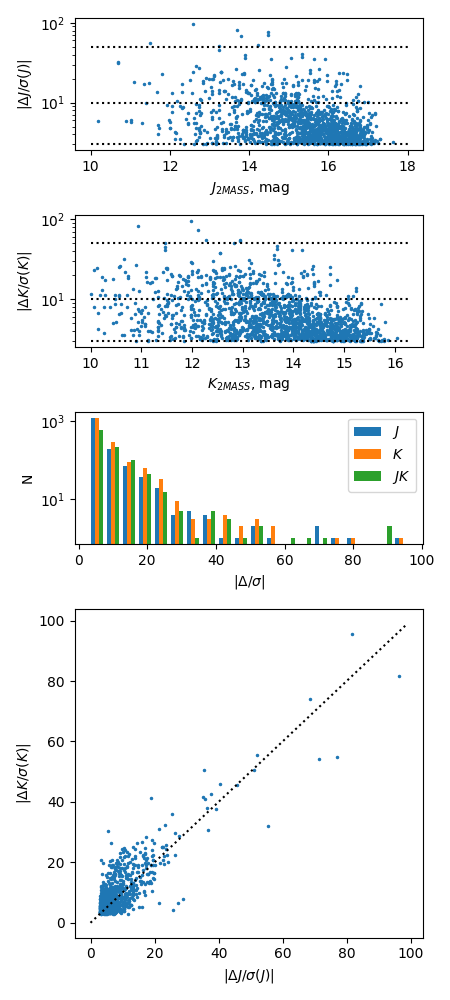}
\caption{Variability properties of quasars in the VCV10 sample. 
The top 
panels show the 2MASS vs. VHS magnitude change in $J$ and $K$ in 
units of measurement errors, added in quadrature from the two 
surveys. The third panel is the histogram of these deviations for 
$J$, $K$ and the average of the two. The bottom panel plots the 
normalized deviations in the two bands against each other; the 
dotted line corresponds to the slope of unity.}\label{fig:sample_VCV10}
\end{figure}

{  Note that here and further throughout the paper the $\sigma$ is
determined based on presumed Poisson photon statistics, which 
ignores any systematic effects that affect the observations, and 
therefore, the quoted significance $\sigma$ levels must be 
interpreted with caution. Furthermore, the variability criterion
is taken at face value, because with archival surveys there is 
no control over the time sampling, so any effects like increasing 
amplitude with time {baseline} can not be accounted.}
The sample that meets all these constraints consists of 1571
objects selected based on $J$ and 1720 selected based on $K$ band
or about 22-24\,\% of the quasars with reliable astrometric match 
between 2MASS and {VHS}.
{952} objects, or 14\,\% are variable in both filters above 3-$\sigma$ 
level and they constitute our final highly variable VCV10 based 
sample, listed in Table\,\ref{tab:var_sample_VCV10} and shown in 
Fig.\,\ref{fig:sample_VCV10}. All these quasars are variable by 
selection, but the histogram (third panel from top to bottom) and 
especially the panel with the $K$ versus $J$ band differences 
(bottom panel) seem to show a valley at above 30-35\,$\sigma$ 
followed by a group of 16 extremely variable objects in both 
filters that will be {given} special attention. {In the subsequent 
analysis, e.g. for the structure function calculation, we consider 
the entire sample, so that the less variable quasars provide a 
baseline that helps us {determine how} the variable quasars 
can be identified from the comparison of these two surveys - 
given their accuracy and cadence.}

We carried out a search for highly variable objects in Quaia
following the same strategy as for the VCV10-based sample -- 
requiring a $>$3$\sigma$ change between 2MASS and VHS observations 
in both $J$ and $K$ bands, to ensure a robust selection. This 
condition was met by 1493 Quaia quasars, listed in 
Table\,\ref{tab:var_sample_Quaua}; 167 of these are {also} 
present in the VCV10-based sample of variable quasars.

\begin{table*}[h!]
\caption{List of 1493 objects, selected from the Quaia sample, 
with photometry from 2MASS and VHS surveys, that show statistically 
significant variability in both $J$ and $K$ bands. Only a fraction 
is shown {as a guide}, the complete table is available in the 
electronic edition of the journal. The table follows the structure 
of Table\,\ref{tab:var_sample_VCV10}. }\label{tab:var_sample_Quaua} 
\tiny
\begin{center}
\begin{tabular}{@{}c@{ }c@{ }c@{ }c@{ }c@{ }c@{ }c@{ }c@{ }c@{ }c@{}} 
\hline
RA DEC (J2000) & 2MASS ID & $J^{2MASS}$ & $K^{2MASS}$ & $J^{VHS}$ & $K^{VHS}$ & ~S/N($\Delta~J$)~ & ~S/N($\Delta~K$)~ & ~S/N($\Delta~J$, $K$)\\
\hline
02:10:46.20 $-$51:01:01.8 &~~J02104620$-$5101018~~~& 13.692 0.026 & 11.977 0.024 & 16.074 0.007 & 14.694 0.009 &  88.25 & 105.76 &  97.01 \\
22:35:13.22 $-$48:35:58.9 & J22351322$-$4835588 & 13.809 0.029 & 12.124 0.026 & 16.018 0.007 & 14.404 0.007 &  74.26 &  84.85 &  79.56 \\
20:05:56.59 $-$23:10:26.9 & J20055658$-$2310269 & 14.489 0.030 & 12.954 0.034 & 17.823 0.035 & 16.331 0.047 &  72.75 &  58.06 &  65.41 \\
\multicolumn{9}{l}{...} \\
20:01:21.44 $-$36:22:59.2 & J20012143$-$3622592 & 15.521 0.061 & 13.665 0.063 & 15.708 0.008 & 13.862 0.006 &   3.04 &   3.12 &   3.08 \\
15:32:02.00 $-$38:47:06.2 & J15320200$-$3847062 & 16.410 0.122 & 14.844 0.124 & 16.787 0.023 & 15.249 0.038 &   3.04 &   3.12 &   3.08 \\
21:20:24.60 $-$47:07:34.3 & J21202460$-$4707342 & 16.880 0.219 & 15.381 0.199 & 17.568 0.032 & 15.992 0.032 &   3.11 &   3.03 &   3.07 \\
\hline
\end{tabular}
\end{center}
\end{table*}

{ 
\subsection{Incompleteness}

Our variability sample is based on two quasar samples that {suffer from} a number of potential problems.
\begin{itemize}
\item The {VCV10 sample} is extremely heterogeneous, because it 
is collected from the literature without regard from the selection 
criteria in the original publications. We can only speculate that 
the incompleteness is an issue mostly for the faint end and the 
bright {objects} are less affected.
\item The Quaia quasar sample is based on the original Gaia 
selection and for quasars it has a relatively low purity of 52\,\%
\citep{2023A&A...674A..41G}. On the other hand, the spectroscopic 
confirmation does improve the sample purity and a comparison with 
independent redshift measurements shows reasonably good agreement 
\citep{2024A&A...687A..16I}.
\item The two photometric samples that we use have different 
completeness limits. \citet{2000AJ....119.2498J} reports that our
first epoch 2MASS is complete down to $J$=15.0\,mag and 
$K_S$=13.5\,mag. Similar numbers are not available in the main 
VHS description paper \citep{2013Msngr.154...35M}, but 
\citet[][figure 5]{2021MNRAS.505.2020E} compares the luminosity 
functions of 2MASS and different VHS sub-surveys, and shows that
VHS -- that provides our second epoch -- is complete to 2.5-3\,mag 
deeper than 2MASS. Of course, this difference depends strongly on
the local crowding. It also implies that our selection would miss 
half of the variable quasars for objects fainter than the 2MASS 
limits quoted above just because we would not consider objects 
that increased their brightness between the two epochs -- they 
would be detected by VSH but {missed by} 2MASS. To 
estimate the incompleteness, we counted the quasars in our VCV10 
based variable sample that become fainter between the two epochs 
and those that became brighter. We ignored the completeness 
limit -- which has some uncertainties itself and varies with the 
stellar density -- and included the entire magnitude range. If 
the two surveys were identical, the two numbers should have been 
equal. Instead, they differ by 78 in $J$ and 66 in $K_S$ bands 
which constitutes $\sim$8\,\% and $\sim$7\,\% of the 952 objects, 
respectively. For the Quaia based sample of variable quasars, the 
differences in the two bands are 925 and 921, or $\sim$19\,\% of
the 1493 objects. We {attribute} this to the fact that VCV10 sample includes 
more bright nearby AGNs than the Quaia sample, and therefore, the 
incompleteness in the former is expected to be smaller than in
the latter.
\item Finally, the quasar variability depends on the time lag, 
and with only two epochs for every object our sample is subject 
to a bias that is an interplay of the varying time baseline in 
the observer's frame and the redshift distribution of the quasars. 
The available data prevent us from investigating how this affects
the completeness of our samples. Most likely this question will 
have to wait for the beginning of the {\it Rubin} operations.
\end{itemize}

Summarizing, our samples of variable quasars {cannot} be considered 
complete and should not be used for statistical applications that 
would rely on completeness, for example, to predict the total 
number of variable quasars down to a given magnitude limit or 
redshift.
}

\section{Ensemble variability}\label{sec:ensable_varisb}

To investigate the collective  variability of the sample we 
performed an ensemble analysis of the quasars in common between 
2MASS and VHS following the framework of \citet{2012ApJ...747...14K} 
who {showed} it is a powerful tool to study the {physical properties
of the nuclear activity}, 
underlying processes and factors driving their variability. In their 
framework the ensemble variability $V$ {  ({i.e.}, the 
magnitude change between the two surveys, averaged over all { objects} 
in the sample)} for a given passband is:
\begin{equation}
V=\sqrt{\frac{\sum_{i}^{N} \Delta m_i^2 - \sum_{i}^{N} \sigma_i^2}{N}}, 
\end{equation}
where $N$ is the number of { objects}, $\Delta m_i$ is the 
magnitude difference and $\sigma_i$ is its observational error
$\sigma_i^2=\sigma_{i, {\rm 2MASS}}^2+\sigma_{i, {\rm VHS}}^2$.
The uncertainty of $V$ is:
\begin{equation}
\sigma_V=\frac{1}{2V} \sqrt{\frac{\left( \sum_{i}^{N} \Delta m_i^2 - \sum_{i}^{N} \sigma_i^2 \right)^2}{N^3}+
\frac{\sum_{i}^{N} \left( 4 \Delta m_i^2 \sigma_i^2 - 2\sigma_i^4 \right)}{N^2}} 
\end{equation}

\begin{figure*}[t!]
\centering
\includegraphics[width=1.7\columnwidth]{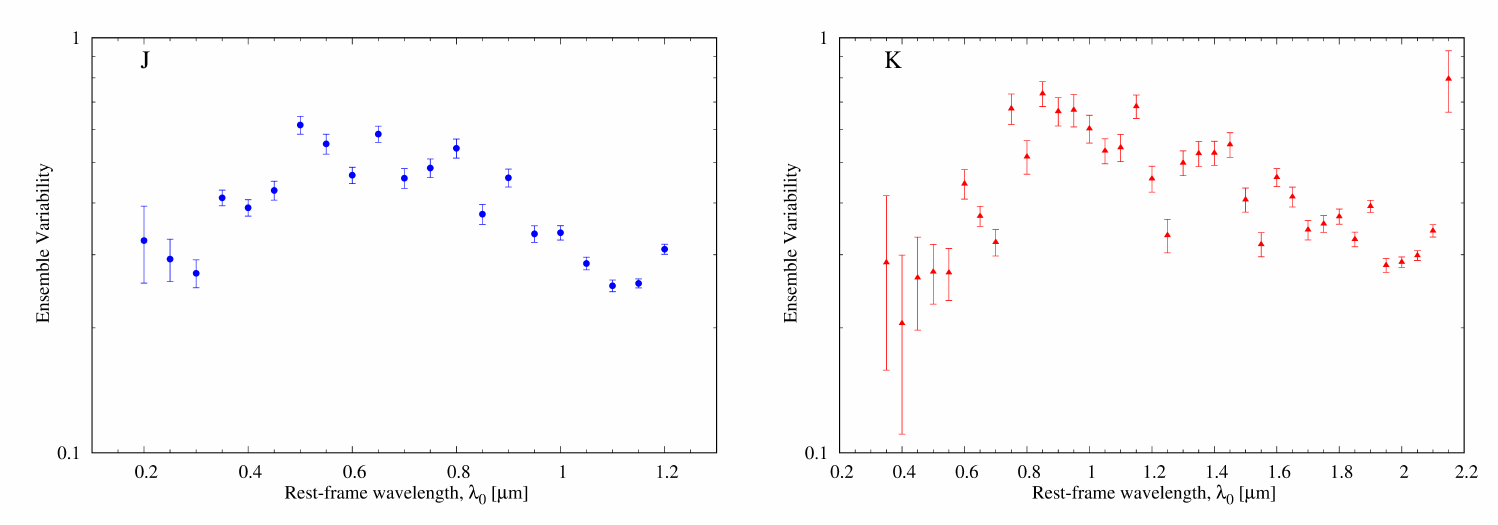}
\includegraphics[width=1.7\columnwidth]{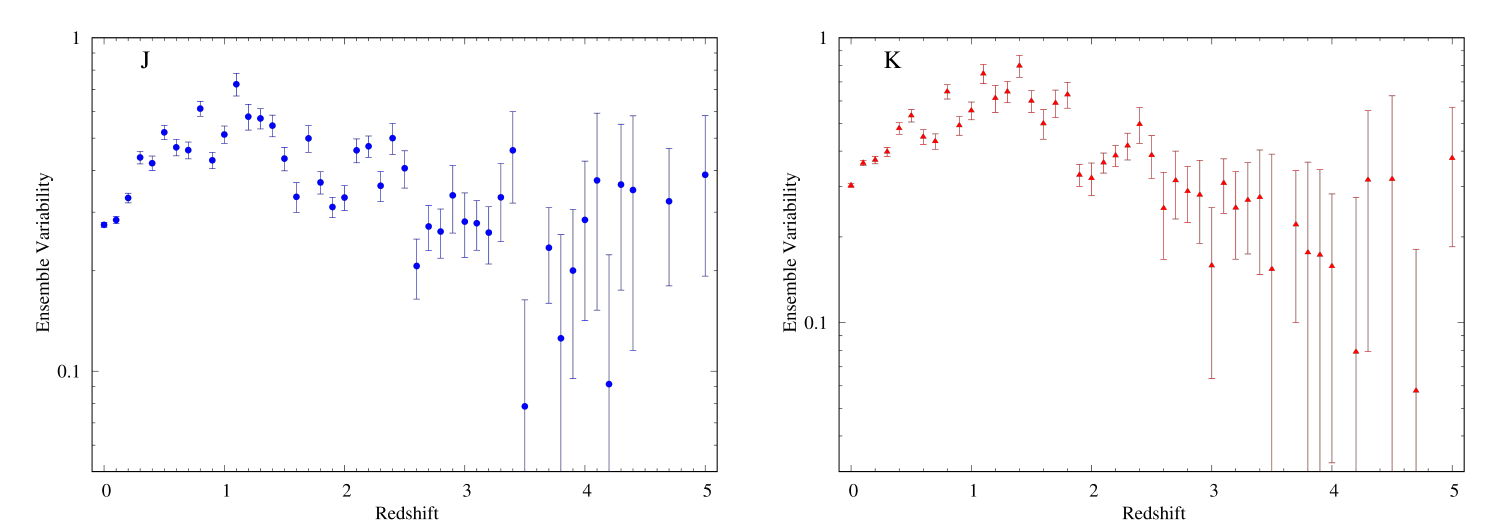}
\includegraphics[width=1.7\columnwidth]{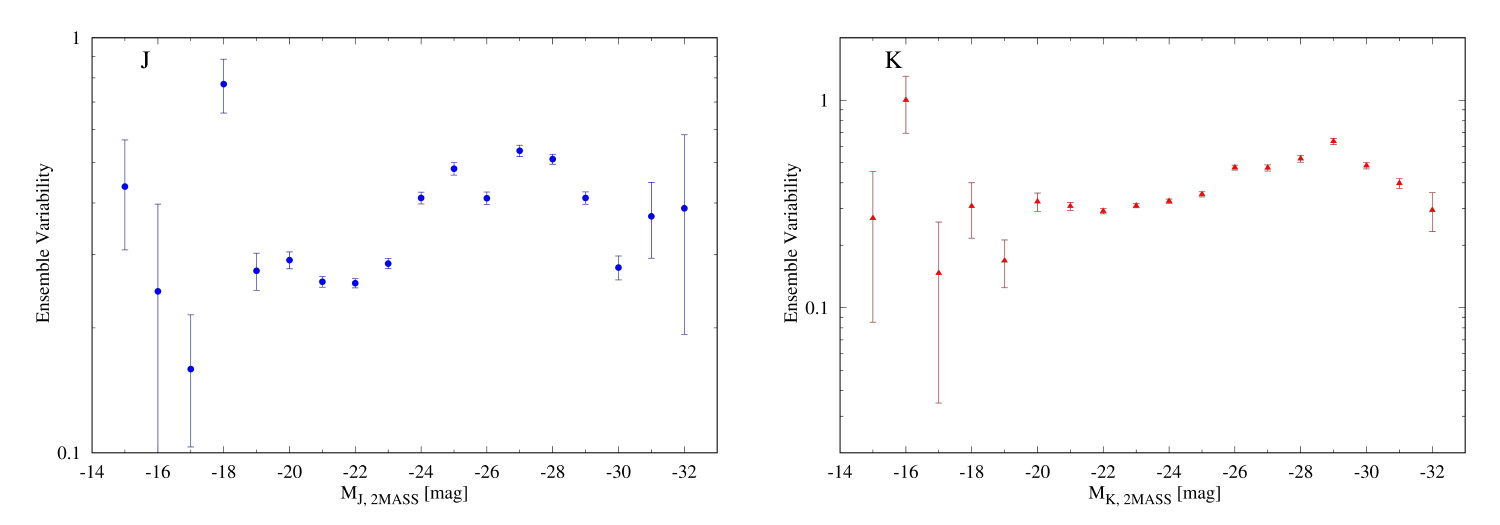}
\caption{Ensemble variability $V$ for the VCV10 sample as a 
function of rest-frame wavelength (top), redshift (middle) and 
near-absolute magnitude (bottom) for the $J$ and $K$ 
band data (left and right, respectively).}\label{EnsVar_RFW_J_K}
\end{figure*}

In the optical (traditionally limited to $\lambda$$\leq$1$\mu$m) 
quasars are known to show weaker variability for longer rest-frame 
wavelength 
\citep{1991ApJ...377..345G,2002ApJ...564..624T,2004ApJ...601..692V,2011A&A...525A..37M}.
Taking advantage of the wide redshift range covered in our sample
we can verify this relation {  (with the disclaimer that only 
10\,\% of our objects have $z$$>$2.0).} Our ensemble variability 
plotted as a function of the rest-frame wavelength is shown in 
Fig.\,\ref{EnsVar_RFW_J_K} (top panels) for both $J$ and $K$ band 
data and we do see a decrease, similar to the relations derived 
by \cite{2012ApJ...747...14K} and shown on their fig.\,4. The 
Spearman 
correlation coefficients are statistically significant -- 0.741 
for $J$ and 0.806 for $K$, {with 99 \% significance} for both 
filters.  The bluest part of the ranges shows an increase and there 
are pronounced maxima at around $\lambda_{rest}$ = 0.5-0.6\,$\mu$m 
for $J$ and at 0.8\,$\mu$m fir $K$. One possibility is that these 
features are due to strong and potentially variable emission lines 
entering the filters' band passes, and contributing to the measured 
flux, for example H$_\beta$ (4861\,$\AA$), [OIII] (5007 and 
4959\,$\AA$) or H$_\alpha$ (6563\,$\AA$). However, the sample 
includes only a few high-$z$ objects, the error bars are larger 
and we refrain from drawing firm conclusions. 

The relations of the ensemble variability with redshift are also 
shown in Fig.\,\ref{EnsVar_RFW_J_K} (middle panels). The quasars 
are more variable with increasing redshift to $z$$\sim$1.2 
{  (this limit {divides} the sample into 80\,\% and 20\,\% fractions)}
and turn down for more distant objects. The reliability of the 
observed trends beyond $z$$\sim$2 is inconclusive due to limited 
statistics. Finally, the ensemble variability as a function of 
absolute magnitudes is plotted in the bottom panels. There are 
peaks at M$\sim$$-$27\,mag and $-$29\,mag. For both relations, we 
observe very similar behaviour to the sample of 
\citet[][their figs.\,5 and 6]{2012ApJ...747...14K}.

\section{Structure functions}\label{sec:sf}

The structure function \citep[SF; ][]{1985ApJ...296...46S} is a 
valuable tool for {parametrising} the nature of variability for 
many objects -- from irregular variable stars to quasars. It 
describes the magnitude change between epochs as a function of 
the time lag between these epochs and the most useful parameter 
is the slope of this relation \citep[see][]{1998ApJ...504..671K}.
The sloped part of the SF is bracketed by two flat regions: the
lower floor is set by the observational errors while the maximum 
is defined by the moments of the most extreme variability. The 
SF has been widely used to study quasars 
\citep{2002MNRAS.329...76H,2004ApJ...601..692V,2005AJ....129..615D, 2009ApJ...696.1241B,2010ApJ...714.1194S,2010ApJ...716..530K,
2012ApJ...747...14K,2016ApJ...817..119K}.

Usually, the SF can be applied to individual objects with 
multi-epoch monitoring that allows to probe a range of time lags. 
Here we apply it on a large sample of objects with two epochs, 
because the time intervals between these observations vary from 
one object to another, so a SF spanning a range of lags can be 
constructed for the sample as a whole. {We caution that the 
surveys we are using have not been optimized to cover well all 
the regimes of the structure function and the linear increase 
section are not clearly traced and the large scatter at in some 
bins makes the choice of the SF fitting range somewhat subjective.}

\begin{figure*}[h!]
\centering
\includegraphics[width=1.7\columnwidth]{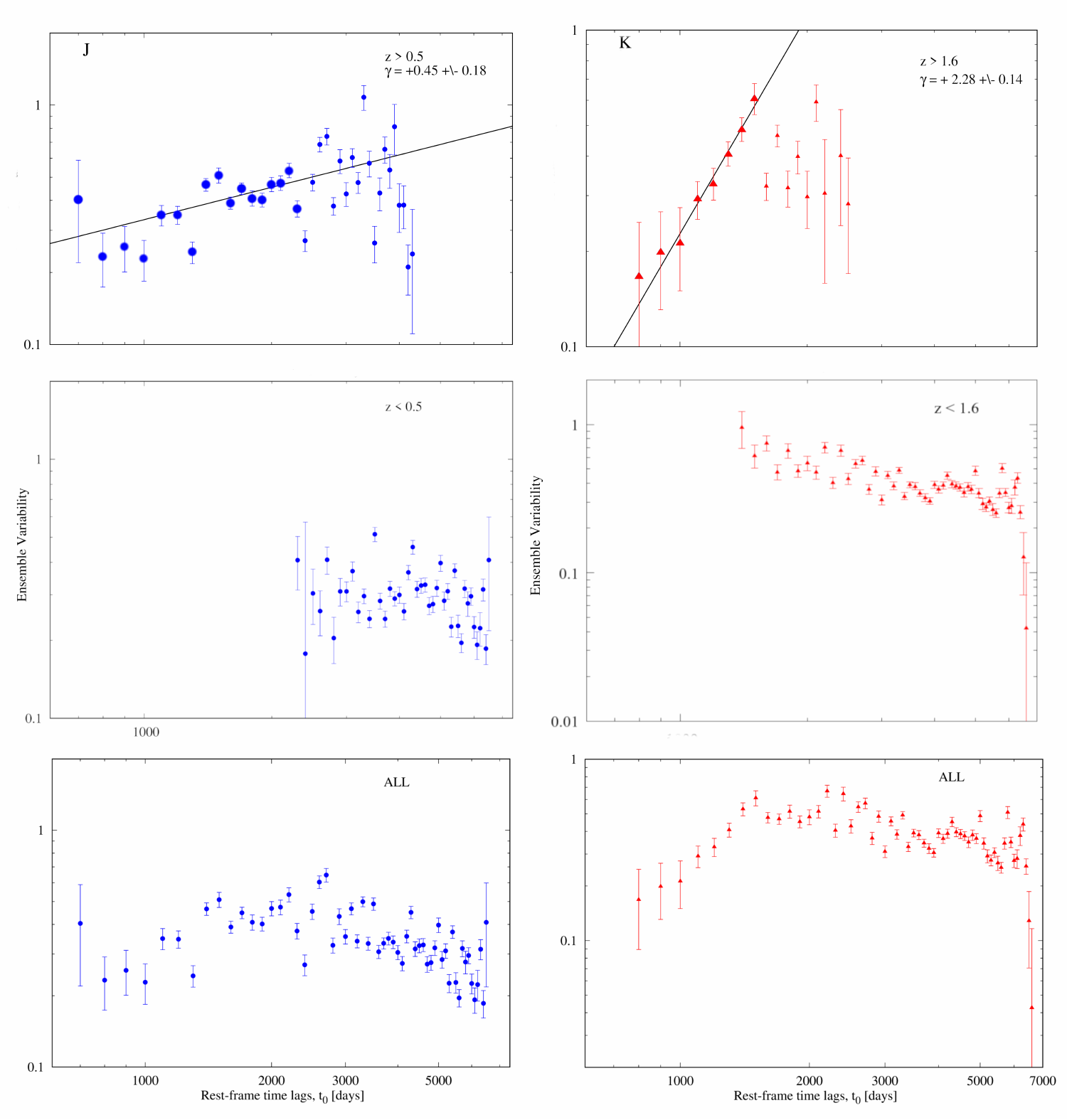}
\caption{Structure functions for $J$ (left) and $K$ (right), 
for redshift-selected subsets and for the entire sample. For 
details see Sec.\,\ref{sec:sf}.}
\label{EnsVar_RFTL_J_K}
\end{figure*}

The SFs for the VCV10-based sample are shown in 
Fig.\,\ref{EnsVar_RFTL_J_K}. {The points used to fit the SF
are marked with larger dots.} We considered all quasars, and 
separately -- nearby and distant subsets, divided at $z$=0.5 
for the $J$ band data and at $z$=1.6 for the $K$ band data. 
The limits were selected to facilitate direct comparison with 
\cite{2012ApJ...747...14K} and we find similar behaviour with
the SFs in their fig.\,7: an increase of the variability for 
rest-frame time lags {below} $\sim$1500\,d and decrease for 
longer lags, giving us confidence in our analysis.

Some of the most relevant theories for explaining the quasar 
variability are the starburst \citep{1992MNRAS.255..713T}, 
instability in the accretion disk \citep{1984ARA&A..22..471R} 
and microlensing. \cite{1998ApJ...504..671K} predicted 
power-law slopes of SF based on various theoretical models. 
In the starburst (supernova) model, the derived SF slope 
$\gamma$ \citep[see eqn. 8 in][]{2012ApJ...747...14K} is in 
the range $\approx$0.7-0.9, compared to the SF slope for the 
accretion disk instability model which results in a slope 
range of $\approx$0.41-0.49. The model based on microlensing 
\citep{2002MNRAS.329...76H} predicted a slope is 
$\approx$0.23–0.31. \cite{2002MNRAS.329...76H} used the 
simulated microlensing light curves of 
\cite{1993MNRAS.261..647L} and \cite{1987A&A...171...49S} 
to generate the SF slopes. 

{
We estimated SF logarithmic slopes $\gamma$ only for the subsets 
of distant quasars: $+$0.45$\pm$0.18 for $J$ ($z$$>$0.5), and 
$+$2.28$\pm$0.14. for $K$ ($z$$>$1.6). The $J$ band slope overlaps 
with the ranges reported in \citet{2012ApJ...747...14K}, but the 
$K$ band slope appears to be excessively large, probably because 
of the ill-defined linear section range.
}

\section{Discussion}\label{sec:discussion}

We look first at the collective properties of the quasars and 
then concentrate on some more prominent individual objects.

\subsection{IR ensemble variability and properties of the quasars 
in other wavelength regimes}

\begin{figure*}[h!]
\centering
\includegraphics[width=1.7\columnwidth]{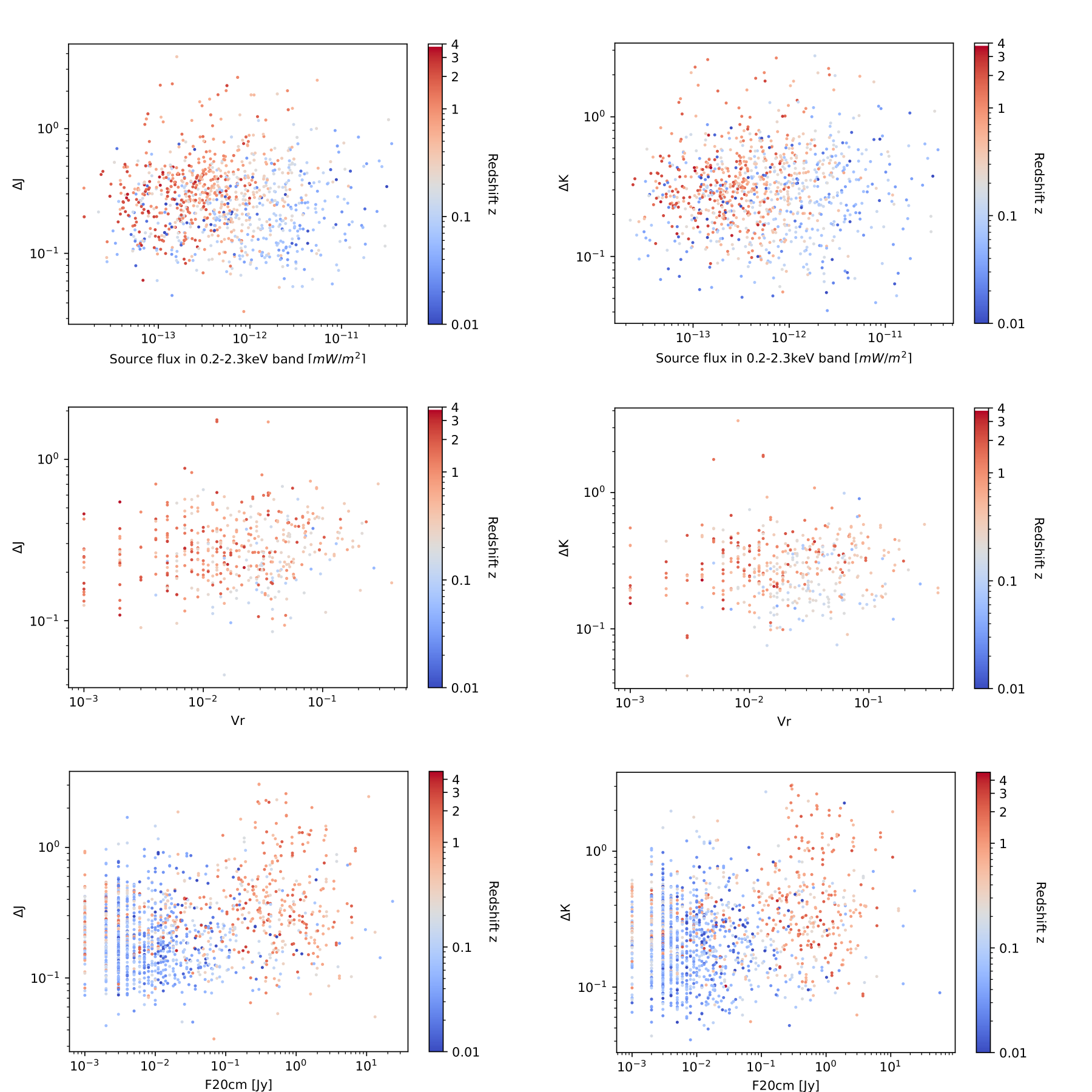}
\caption{IR variability amplitude in $J$ (left) and $K$ (right) 
versus X-ray luminosity (top panels), optical variability in the
SDSS $r$ band (middle panels) and radio luminosity (bottom 
panels). The points are coloured by redshift, according to the 
respective colour bars.}\label{Radio_vs_NIR}
\end{figure*}

There is a discussion of the radio emission and the quasar 
optical variability {correlation} 
\citep[e.g.,][]{1983ApJ...272...11P,2009ApJ...696.1241B}. 
\cite{2012ApJ...747...14K} reported a weak increase {in} the IR 
variability with increasing radio luminosity. To verify this 
result we collected 20\,cm luminosities for the quasars in the 
VCV10 sample from Faint Images of the Radio Sky at Twenty 
Centimeters \citep[FIRST;][]{1995ApJ...450..559B} and we 
confirm this {trend} (Fig.\,\ref{Radio_vs_NIR}, bottom panels). 
{  Indeed, the Spearman test yields a positive non-zero coefficient of 
0.22 for filter $J$ (p-value < 0.001, 95\% confidence interval: 0.16 to 
0.27) and 0.20 for filter $K$ (p-value$<$0.001, 95\% confidence 
interval: 0.14 to 0.24). These results suggest a statistically 
significant, though weak, positive correlation between radio luminosity 
and IR variability in both the $J$ and $K$ filters.}

Next, we compare the IR variability of the objects in the VCV10 
sample with the UV/optical ensemble variability estimated by 
\citet{2011A&A...525A..37M} in {the} $r$ filter from the 
Sloan Digital Sky Survey \citep[SDSS;][]{2000AJ....120.1579Y}. 
The results are shown in Fig.\,\ref{Radio_vs_NIR} (middle panels). 
\cite{2012ApJ...747...14K} claim a weak correlation for $J$ band 
and weak anti-correlation for $K$.
{The Spearman test for our data yields a coefficient of 0.14 for 
the $J$ band (p-value = 0.003, 95\% confidence interval: 0.05 to 
0.22), indicating a weak but statistically significant positive 
correlation. For the $K$ band, the coefficient is 0.06 
(p-value = 0.24, 95\% confidence interval: $-$0.04 to 0.15), 
suggesting virtually no correlation. Our $J$ band correlation is 
slightly stronger than that reported by \citet{2012ApJ...747...14K}, 
but for $K$, we find negligible evidence of a relationship.} 
A larger data set is needed to draw firmer 
{conclusions}, but both our analysis and that of 
\citet{2011A&A...525A..37M} hint -- not surprisingly -- that 
quasars may behave more similarly in filters with closer 
central wavelengths than in filters with more separate 
wavelengths.

Finally, we compare the magnitude changes of VCV10 quasars 
with their X-ray luminosities at 0.2-2.3\,keV from the 
\citet{2024A&A...682A..34M} {catalogue}, based on observations 
with {the} extended ROentgen Survey with an Imaging Telescope 
Array \citep[eROSITA;][]{2020Natur.588..227P}, on board the 
\citep[SRG;][]{2021A&A...656A.132S} orbital observatory. { For the 
$J$ band, we find essentially no correlation, with a Spearman 
coefficient of $-$0.00008 (p-value = 0.998, 95\% confidence interval: 
$-$0.06 to 0.06). For the $K$ band, we find a very weak positive 
correlation, with a coefficient of 0.07 (p-value = 0.041, 95\% 
confidence interval: $-$0.004 to 0.12). These results indicate minimal 
evidence of correlation between X-ray luminosity and IR variability, 
with only a slight association observed in the $K$ band} 
(Fig.\,\ref{Radio_vs_NIR} bottom panels).

\subsection{Nature of the most variable quasars}

\begin{figure*}[h!]
\centering
\includegraphics[width=1.7\columnwidth]{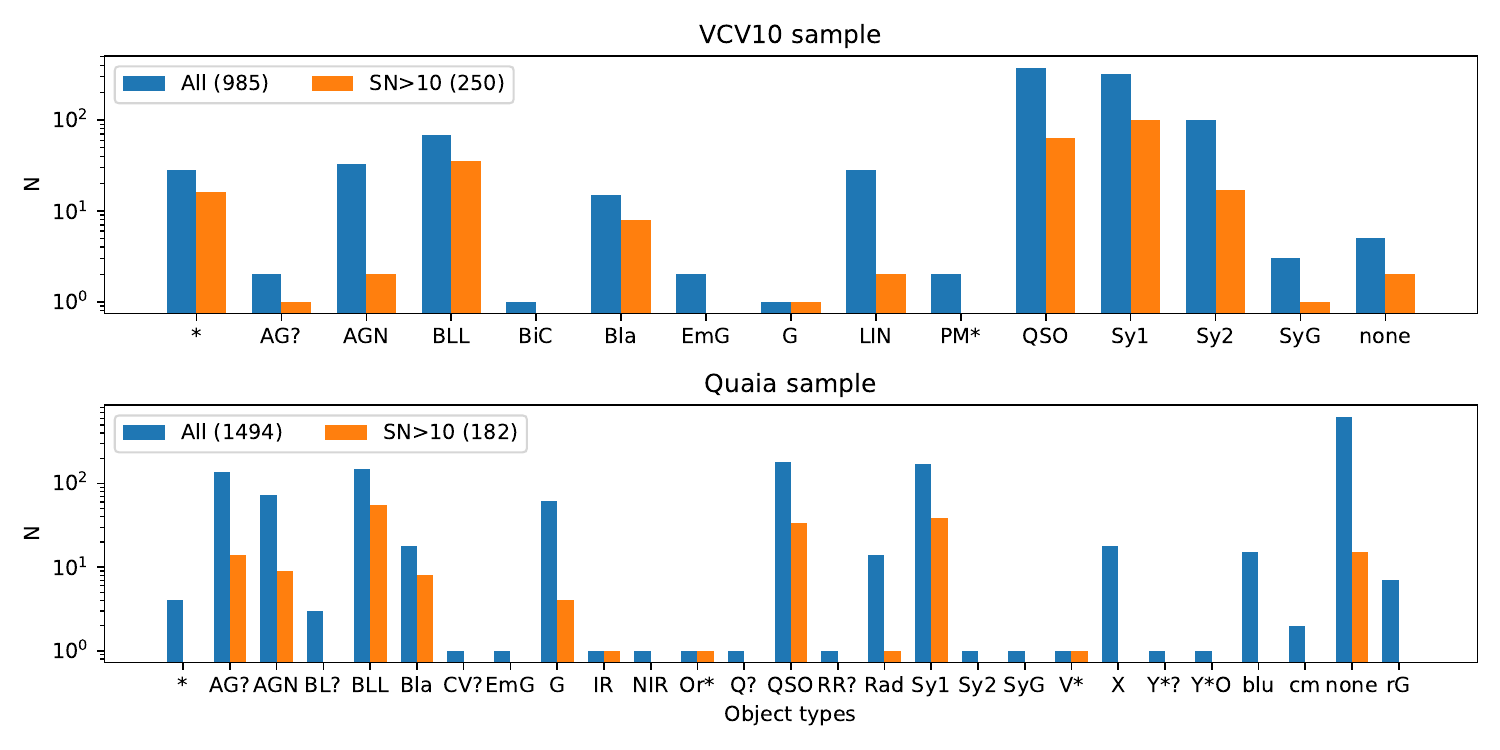}
\includegraphics[width=1.7\columnwidth]{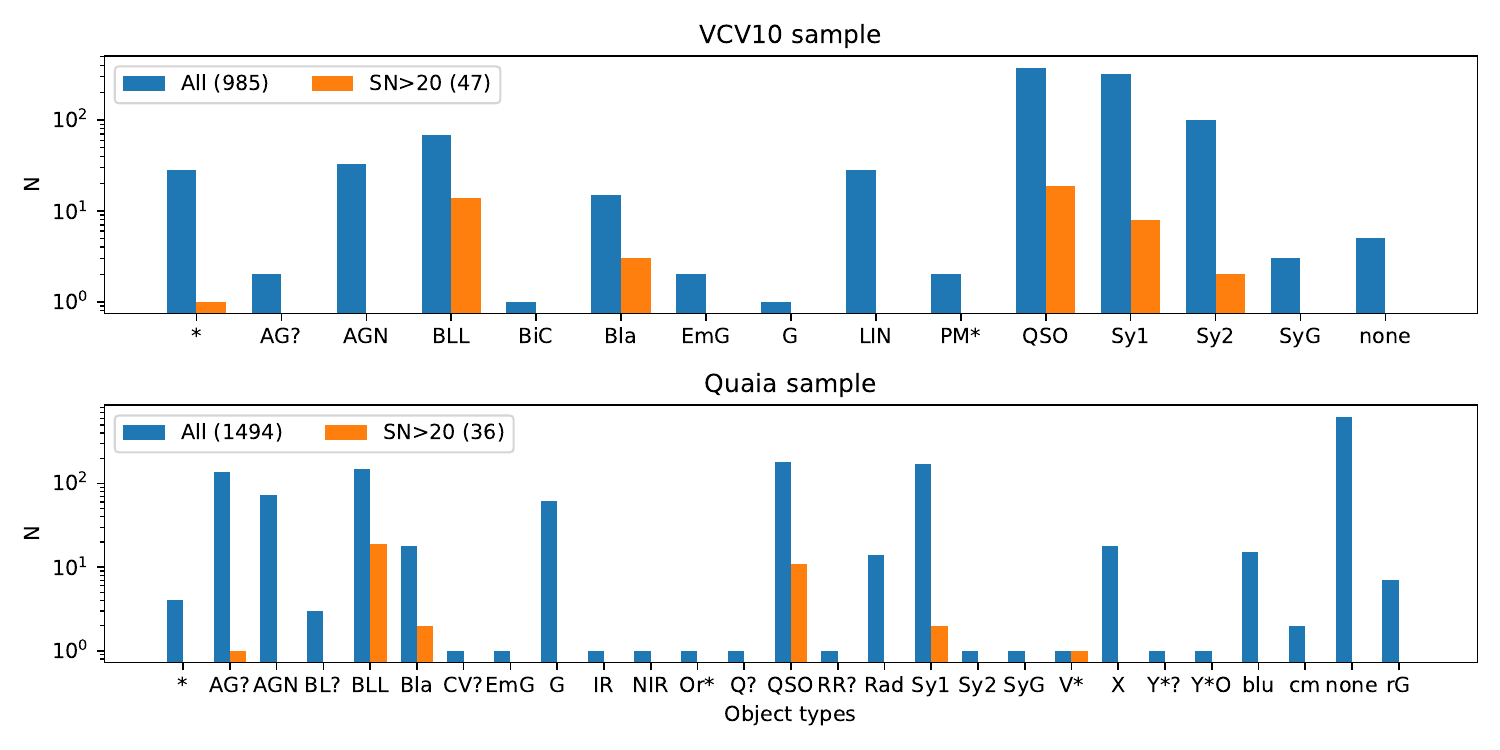}
\caption{Distributions of SIMBAD object types for the VCV10 
and Quaia (top and bottom, alternating) samples for different 
variability levels: the blue histograms show the complete 
samples and {the orange histograms} show extremely variable objects with 
average changes in $J$ and $K$ that exceed 10- and 20-$\sigma$ 
significance {levels} (upper and lower pair of panels). Bracketed 
numbers in the legends show the number of objects in each set. 
The bin for objects with no SIMBAD classification {is} marked 
with {"none"}.}\label{fig:hist_OType}
\end{figure*}

We looked at the nature of the most variable objects in the 
{VCV10-based} sample, because the original VCV10 sample consists 
mostly of well-known active galaxies that have been studied 
for a long time and they can {provide} a more definitive picture of 
the selected objects. We collected their SIMBAD 
types\footnote{Information about the SIMBAD types and their 
hierarchical structure is available at 
\href{http://vizier.u-strasbg.fr/cgi-bin/OType} {http://vizier.u-strasbg.fr/cgi-bin/OType}.}, 
with a clear understanding that this classification is based
on heterogeneous data and analysis. Many objects have 
multiple matches within 2\,arcsec with different types, but 
in most cases, these correspond to detections of the same {object} at
different wavelength regimes. For example, if an object is a 
QSO and there are nearby sources in X-ray or UV or Radio -- we 
retained the sole QSO type. The distribution of the 852 most 
variable objects in the VCV10-based sample by type is shown 
in Fig.\,\ref{fig:hist_OType}.

{  
A check in SIMBAD indicates that objects with classifications 
among both samples of variable objects (based on VCV10 and 
Quaia -- Tables\,\ref{tab:var_sample_VCV10} and 
\ref{tab:var_sample_Quaua}, respectively)} are dominated by 
quasars, Seyferts, BL\,Lacs or {blazars}, but the Quaia sample 
includes a host of other objects: some of them have been 
misclassified in the older literature as various types of 
stars (young stellar objects, various types of variable stars, 
CVs, etc.) and others are only known by the wavelength range 
of their detection (IR, radio, etc.). Remarkably, {  there are
626 objects in the Quaia sample  with no classifications 
(marked as none in Fig.\,\ref{fig:hist_OType}) and they 
constitute} the single most populated class, underlying the 
importance of further studies of this new quasar sample.
{  Imposing a requirement for a variability amplitude between 
the 2MASS and VHS epochs of ten or twenty times the typical 
observing errors (top and bottom pairs of panels, respectively) 
removes the majority of {objects} that were classified in SIMBAD 
as ``stellar''} and leaves subsets of objects that have been 
classified as quasars, BL\,Lacs or Seyfert galaxies.

The most variable in the Quasia sample is the well-known 
$z$=1.033 flat spectrum radio quasar (FSRQ) QSO\,B0208$-$512. 
It already has been subjected to rigorous optical spectral 
monitoring that indicated the redder part of the spectrum 
is more stable when the object is brighter than the bluer
part \citep{2022ApJS..259...49Z}. It is followed by the 
$z$=0.506 FSRQ QSO\,B2232$-$488 that shows frequently seen 
redder-when-brighter behaviour in the optical 
\citep{2015RAA....15.1784Z} and the $z$=0.833 blazar
[VV2006]\,J200556.6$-$231027 \citep{2002ApJ...569...36W}.
The latter object has also been monitored in the optical 
for variability by \citet{2021AJ....162...21B} who find 
that it showed an amplitude of 1.230\,mag in $g$ band, and 
that the amplitude {decreased} with wavelength; the smallest 
amplitude was 0.386\,mag in $y$. This is different from the 
dominant {behaviour} in their sample of 2863 quasars that 
typically exhibit the lowest variability in $r$ band and 
the largest -- in $y$ band. This contrast hints that our
selection does indeed find unusual objects.

To assess broadly how {recognised} are members of our Quaia 
based sample of variable quasars we searched for them in 
SIMBAD and found that 587 or $\sim$39\,\% have no known 
counterparts within a rather large search radius of 
10\,arcsec, especially when the average separation for the 
identified counterparts is 0.6\,arcsec. Out of the remaining 
908 quasars 526 have ten or fewer references -- they have 
hardly been studied. One such example is the $z$=0.08513 
Seyfert 1 Galaxy 2MASS\,J21572316$-$3123064 -- object 43 
in Fig.\,\ref{fig:ZTH_LCs}, with extremely variable 
behaviour in the optical (see the next section).

\subsection{Optical light curves}

\begin{figure}[t!]
\centering
\includegraphics[width=1.0\columnwidth]{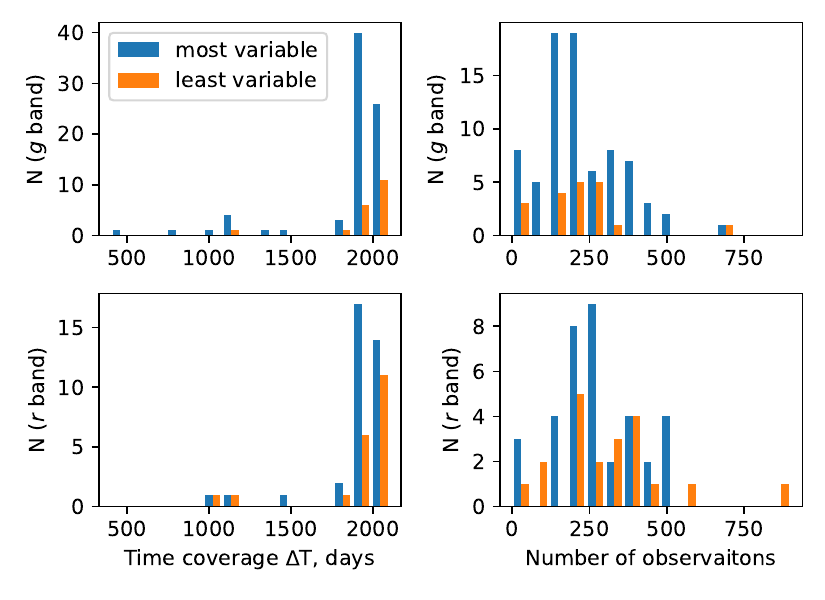}
\caption{Properties of the optical ZTF light curves for the 
most variable 80 (blue) and for the least variable 20 objects
(orange) in our sample. Only optical light curves with at 
least 3 observations are included here, {which reduces} these 
numbers. 
{\it Left:} Histograms of the time coverage of the light 
curves for $g$ (top) and $r$ (bottom) filters. 
{\it Right:} The same plots for the number of observations.
}\label{fig:opt_LCs}
\end{figure}

\begin{figure}[h!]
\centering
\includegraphics[width=1.0\columnwidth]{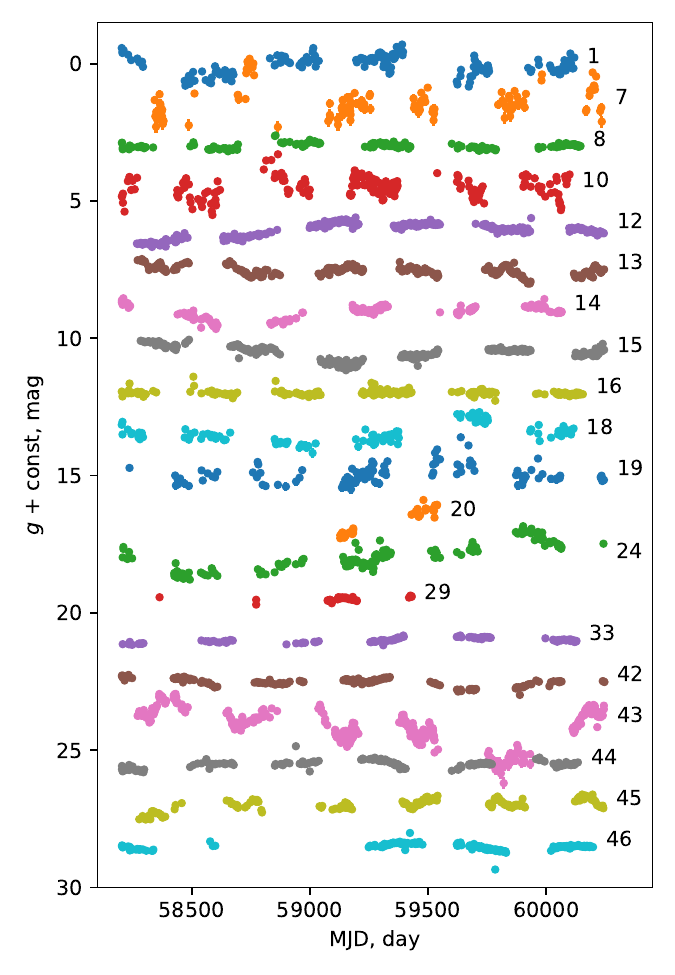}
\caption{ZTF $g$ band light curves of a subset of the most 
variable Quaia quasars, according to the 2MASS versus VHS 
magnitude differences. The number on the right corresponds to 
the sequential number in Table\,\ref{tab:var_sample_Quaua}. 
The observational error bars are smaller than the 
symbols.}\label{fig:ZTH_LCs}
\end{figure}

About 40\,\% of the objects in our selection of extremely 
variable quasars have been observed with ZTF {  (we only 
consider objects with at least 3 observations). The number
of epochs usually varies from $\sim$30 to 890, with an 
average of 255; the covered time span is between 410 days
and 2050 days, {with} most objects ($\sim$85\,\% for $g$ and
$\sim$90\,\% {for} $r$) well above 1800 days 
(Fig.\,\ref{fig:opt_LCs}).} Some 
$g$ band light curves are shown in Fig.\,\ref{fig:ZTH_LCs}
with the more variable objects in the near-IR on the top
and not surprisingly, they tend to show larger variation in 
the optical as well. However, there are {exceptions}, like 
objects 8 and 16 that show nearly flat light curves.

\begin{figure}
\centering
\includegraphics[width=1.0\columnwidth]{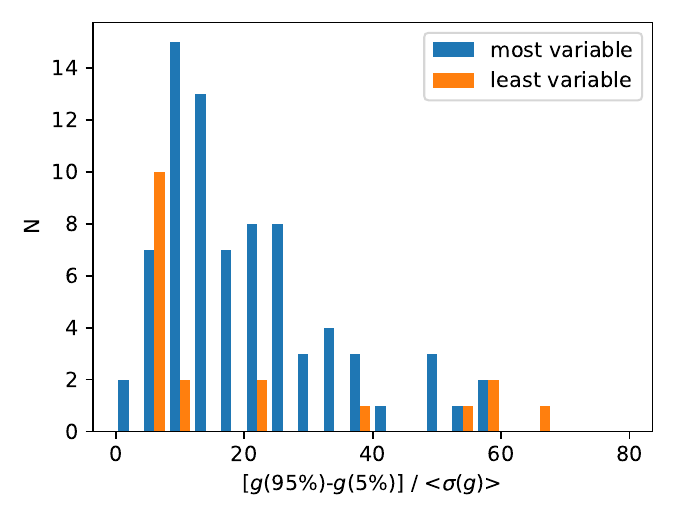}
\caption{Histograms of the 5 and 95 percentile magnitudes 
for the ZTF light curves of the most and least variable 
candidates from Table\,\ref{tab:var_sample_Quaua}}\label{fig:opt_LC_ampl}
\end{figure}

Our infrared selection is based only on two epochs, so it
includes highly variable objects that may have undergone 
an episodic or short-lived change. We {cannot} apply a more 
robust variability indicator for the 2MASS and VHS data 
sets, but given the significant number of epochs in the ZTF, 
we can do this for that data set. To exclude outliers and to 
limit our considerations to objects that do show consistently 
variable behaviour over long term -- which makes them better 
suited for reverberation mapping, for example -- we measured 
the optical amplitude of each light curve as a difference 
between the 5\,\% and the 95\,\% magnitudes. Furthermore, we 
{normalised} this difference by the median observational error 
for all the observations in that light curve. The histograms 
of the {normalised} amplitudes for the 80 most variable objects
with available ZTF data, and for the 20 least variable in 
Table\,\ref{tab:var_sample_Quaua} are shown in 
Fig.\,\ref{fig:opt_LC_ampl}. As expected, the former subset 
shows systematically larger amplitudes than the latter. 

If we tentatively assume that the objects in the first two 
bins on the left of the histogram are non-variable -- an 
overly pessimistic assumption, because all of them fall well 
above the 3-$\sigma$ variability limit by sample construction -- 
then our 2-epoch variability criterion allows for $\sim$11\,\% 
of non-variable quasar contamination near the {least} variable end 
of our ranked sample. This is a strikingly high efficiency for 
the selection of variable quasars, given the vast difference in 
the number of epochs -- two for our selection (albeit in two 
photometric bands) versus the typical number of over two hundred 
for the ZTF.

\section{Summary and conclusions}\label{sec:conclusions}

We selected a sample of nearly 2.5 thousand variable quasars 
for future variability and {reverberation} studies from the 
historic VCV10 {catalogue} and from the recent Quaia quasar 
catalogue, which identified quasars based on Gaia spectra and 
mid-infrared colours. Our selection is based on the difference 
{between} near-IR $J$ and $K$ band observations of 2MASS and VHS,
normalized by the observational errors and averaged over the
two bands. These two surveys are separated by nearly a decade
which ensures that most of the variable objects exhibit 
consistently variable behaviour over long time scales. 

To verify that 2MASS and VHS surveys cover a representative 
sample of quasars we applied an ensemble variability and 
structure function {analysis} and compared our results with 
the work of \citet{2012ApJ...747...14K}, finding good 
agreement. This strengthened some conclusions about the 
quasars' ensemble variability, for example that it decreases 
with the rest-frame wavelength or that it exhibits peaks for 
certain absolute magnitudes of the quasars. Similarly, the 
structure function shows an increase in the variability for 
rest-frame time lags {below} $\sim$1500\,d and decrease for 
longer lags, just like in previous studies. {  {Their} trends 
agree qualitatively with the ones found in optical long-term
studies of quasar variability 
\citep{2003AJ....126.1217D,2005AJ....129..615D}.
}

Aware of the limited number of observations we used, we 
investigated if the comparison of only two epochs could 
provide a reliable selection of variable quasars. 
Our Quaia based sample of variable quasars consists of 1493
objects and $\sim$40\,\% of them have ZTF optical light 
curves. We inspected a subset of these, finding that for 
the strongest candidates our method allows for $\sim$11\,\% 
of contamination by quasars that show only weak optical 
variability. 
Despite this contamination, we find it remarkably efficient 
that the comparison of only two epochs in two filters yields 
results consistent with the analysis of light curves with 
two orders of magnitude more measurements. {  A brief inspection
of the literature indicated that the vast majority of our
Quaua based variable quasar sample has not been studied in 
{detail} before -- they are either completely absent from 
SIMBAD or are only present as entries in various survey 
catalogs, increasing the value of this sample for further
quasar variability studies.}

A final word of caution: our samples of variable quasars suffer 
from a number of {incompleteness issues} and should not be 
used for statistical studies.

...........................................................

\begin{acknowledgements}
This research has made use of NASA’s Astrophysics Data System 
Bibliographic Services. 
This research has made use of the VizieR catalogue access tool, 
CDS, Strasbourg, France.
Topcat \citep{2005ASPC..347...29T}, 
Astropy \citep{2013A&A...558A..33A,2018AJ....156..123A} and
Matplotlib \citep{Hunter:2007}
tools were used in this research.
We acknowledge support by Bulgarian National Science Fund under 
grant DN18-10/2017 and Bulgarian National Roadmap for Research 
Infrastructure Project D01-326/04.12.2023 of the  Ministry of Education 
and Science of the Republic of Bulgaria.
\end{acknowledgements}


\bibliographystyle{aa}
\bibliography{bibliography_paper.bib} 

\end{document}